\documentclass[aps,preprintnumbers,eqsecnum,amsmath,amssymb,nofootinbib]{revtex4}
\usepackage{graphicx}
\usepackage[english]{babel}
\usepackage{url}
\usepackage{graphicx}
\usepackage{xcolor}
\usepackage{amsmath}
\usepackage{amssymb}
\usepackage{slashed}
\usepackage{multirow}

\begin{document}
\title{Factorization of multiparticle contributions to amplitudes of $B$-meson weak decays}
\author{Dmitri Melikhov$^{a,b,c}$}
\affiliation{
$^a$D.~V.~Skobeltsyn Institute of Nuclear Physics, M.~V.~Lomonosov Moscow State University, 119991, Moscow, Russia\\
$^b$Joint Institute for Nuclear Research, 141980 Dubna, Russia\\
$^c$Faculty of Physics, University of Vienna, Boltzmanngasse 5, A-1090 Vienna, Austria}
\date{\today}
\begin{abstract}
We show that multiparticle contributions to amplitudes of weak decays of 
the generic topology (heavy quark hits some intermediate point of the propagator line joining the end-points from
which momenta $q$ and $q'$ are emitted) is given in the heavy quark limit and at the leading order in $\alpha_s$ 
by the convolution of (i) hard kernel composed of highly virtual 
propagators of light degrees of freedom and (ii) the $B$-meson multiparticle wave function,
$\langle 0|\phi(x)\phi(x_1)\dots\phi(x_n)\phi_b(0)\phi(x'_{n'})\dots\phi(x'_1)\phi(x')|B(p)\rangle$, 
in a {\it double-collinear} light-cone configuration: the coordinates $x,x_1,...,x_n$ are
ordered and aligned along the light-like 4-vector $a_\mu$, $a^2=0$, $q_\mu\propto a_\mu$, 
while the coordinates $x',x_1',...,x'_{n'}$ are ordered and aligned along the light-like 4-vector $a'_\mu$,
$a'^2=0$, $q'_\mu\propto a'_\mu$, $a'a\ne 0$. Corrections to this factorization formula are suppressed by
powers of $\Lambda_{\rm QCD}/M_B$. 
\end{abstract}
\maketitle


\section{\label{Sec:1}Introduction}
\begin{figure}[b]
  \begin{center}
\includegraphics[height=6cm]{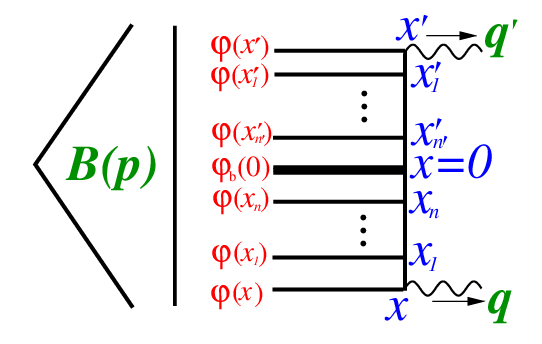}
  \caption{\label{Fig:0a}
    A diagram of the generic form factor topology: the heavy $b$-quark field which we place at $x=0$
    hits the intermediate point of the propagator line between $x$ and $x'$.
    The part of the propagator line between $x=0$ and $x$,
    at which the momentum $q$ is emitted, contains $n$ intermediate points coupled to the
    light fields $\phi(x_1)...\phi(x_n)$.
    The part of the propagator line between $x=0$ and $x'$, at which
    the momentum $q'$ is emitted, contains $n'$ intermediate points coupled to the
    light fields $\phi(x'_n)...\phi(x'_1)$. The corresponding amplitude is denoted as $A^{(n,n')}(q,q')$.}
\end{center}
\end{figure}

In this paper we give a proof of factorization formula for multiparticle contributions
to amplitudes of heavy $B$-meson decays
of the generic weak form factor topology: the heavy-quark $b$ field hits some {\it intermediate point} on the
line along which energetic light degrees of freedom propagate in the Feynman diagram.
Fig.~\ref{Fig:0a} shows $A^{(n,n')}(q,q')$, a typical example of an amplitude of our interest.

\begin{figure}[t]
  \begin{center}
\includegraphics[height=5cm]{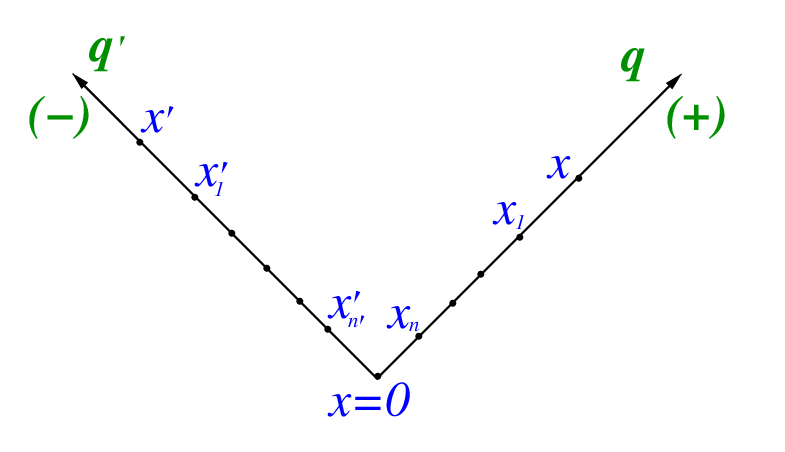}
  \caption{\label{Fig:0b}
    The double-collinear light-cone field configuration of the wave function of Eq.~(\ref{BSresult}): 
    The field coordinates $x,x_1,...,x_n,0$ are ordered along the $(+)$ axis of the light cone; 
    The field coordinates $x',x'_1,...,x'_{n'},0$ are ordered along the $(-)$ axis of the light cone.
    In the reference frame where $q=(q_+,0,0)$ and $q'=(0,q'_-,0)$, this very field configuration
    gives the dominant contribution to the amplitude of the generic weak form factor topology.}
\end{center}
\end{figure}

We are going to demonstrate that at the leading order in $\alpha_s$, 
the amplitude $A^{(n,n')}(q,q')$ satisfies the following factorization
formula in the limit $m_b\to\infty$:
In the reference frame where momentum $q$ has a large $(+)$ component and the momentum $q'$
has a large $(-)$ component, $A^{(n,n')}(q,q')$ is given by the convolution integral of
(i) the hard kernel composed of propagators
of light degrees of freedom and (ii) the $B$-meson multiparticle wave function in the following
{\it double collinear light-cone configuration} shown in Fig.~\ref{Fig:0b}:
\begin{eqnarray}
  \label{BSresult}
  &&\langle 0|\phi(x_+)\phi(u_1 x_+)...\phi(u_nx_+)
  \phi_b(0)
  \phi(u'_{n'}x'_{-})\dots\phi(u'_1x'_-)\phi(x'_-)|B(p)\rangle,\nonumber\\
  &&\qquad\qquad 0<u_{n}<...<u_1<1,\quad 0<u'_{n'}<...<u'_1<1. 
\end{eqnarray}
Here the heavy field $\phi_b$ ($b$-quark) and the light fields $\phi(x),....\phi(x')$ (light quarks or gluons)
are understood as Heisenberg operators. Eq.~(\ref{BSresult}) means that the coordinates of the field operators in the lower [upper]
block of the diagram of Fig.~\ref{Fig:0a} are ordered and aligned along the $(+)$ [$(-)$] light-cone (LC) direction,
i.e., along the direction of $q$ [$q'$]. 

Well-known weak semileptonic (SL) form factors of the $B$-meson represent a degenerate case of the generic weak
form factor topology and correspond to the situation when the heavy-quark field hits
the {\it end-point} of the propagator line mentioned above. As the result (see, e.g., \cite{offen2007,wang2022a})
three-particle contribution to $B$-meson SL form factors may be expressed 
via the quark-antiquark-gluon three-particle $B$-meson wave function in a {\it collinear LC configuration}, 
$\langle 0|\bar s(x) G_{\mu\nu}(u x)b(0)|B_s(p)\rangle$, $x^2=0$, $0<u<1$ \cite{japan2001}.
Our analysis of Sec.~\ref{Sec:3} gives a generalization of this result to an arbitrary number
of the emitted soft gluons. 

Charming loops in flavour-changing neutral current (FCNC) $B$-meson decays \cite{Neubert}
represent an example of the generic form factor topology \cite{m2022}.
Charming-loop amplitudes involving 3-particle quark-antiquark-gluon contributions were studied
in \cite{voloshin,ligeti,buchalla,khod1997,zwicky1,zwicky2}. 
In \cite{hidr}, followed by \cite{gubernari2020,gubernari2022}, it was asserted that the 
contribution of charming loops to the amplitudes of $B\to (K,K^*)l^+l^-$ decays
may be calculated via the same collinear LC 3-particle wave function
as in the SL form factor.
This result was put under scrutiny in \cite{mk2018,m2019,m2023} where it was found
that the factorization formula for charming loops in FCNC $B$-decay contains 
three-particle wave function in a different -- a {\it noncollinear LC configuration}: 
$\langle 0|\bar s(x) G_{\mu\nu}(x')b(0)|B_s(p)\rangle$, $x^2=0$, $x'^2=0$, but $x'x\ne 0$.
The applications of this noncollinear LC wave function to FCNC $B$-decays in QCD was presented
in \cite{wang2022,wang2023,bbm2023,bbm2024}. 

This paper gives a proof of the factorization formula for the case of muliparticle contributions to weak $B$-decays
of the generic topology. For the sake of technical simplicity and clarity, all discussions are presented
for scalar particles, still keeping the notation ``quarks'' and ``gluons''.
Complications of QCD related to gauge invariance and quark and gluon spins do not change the essential
features of our analysis.
Moreover, we use the notation $\Lambda_{\rm QCD}$ as a typical hadron scale of a few hundred MeV, 
$\Lambda_{\rm QCD}\ll m_b$.

In Sec.~\ref{Sec:2} we discuss the 3-particle contribution to the amplitude of the generic $B$-decay topology.
Sec.~\ref{Sec:3} obtains factorization formula for multiparticle contribution to the amplitude of SL topology.
Finally, Sec.~\ref{Sec:4} presents factorization formula for multiparticle contribution to the amplitude of
the generic weak decay topology. Sec.~\ref{Sec:5} gives a brief discussion of the radiative corrections and 
Sec.~\ref{Sec:6} gives our summary. 

\section{\label{Sec:2} 3-particle contribution to the amplitude of the generic weak form factor topology}
\begin{figure}[b]
  \begin{center}
    \includegraphics[height=5cm]{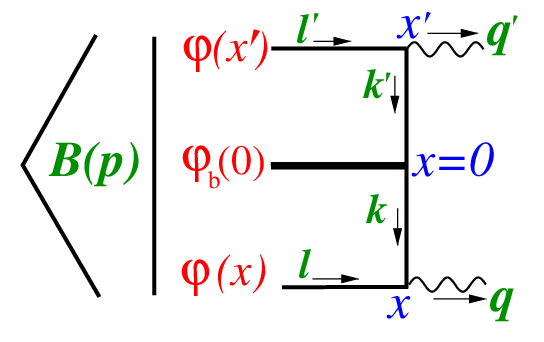}
    \caption{\label{Fig:1}Three-particle contribution to the amplitude of the generic weak form factor topology:
      the heavy field $\phi_b$ hits the intermediate point of the (vertical) line along which light degrees of freedom propagate.
    Except for $\phi_b$, all other fields in the multiparticle wave function are light fields.}
    \end{center}
\end{figure}

In this Section, we obtain factorization formula for the simplest amplitude of the generic weak form factor topology of
Fig.~\ref{Fig:1}, where the heavy field hits the intermediate point of the propagator line along which light degrees of
freedom propagate. This amplitude involves the matrix element of three field operators and reads:  
\begin{eqnarray}
  \label{A0}
  A(q,q')=\int dx dx' dk dk'\, D(k)\,D(k')\,e^{i q x+i k x +i q' x'-i k' x'}\Psi(x,0,x'|p),
  \quad p=q+q',
\end{eqnarray}
where  
\begin{eqnarray}
  \label{3BS}
  \Psi(x,0,x'|p)= \langle 0|\phi(x)\phi_b(0)\phi(x')|B(p)\rangle 
\end{eqnarray}
is the 3-particle wave function of the $B$-meson. The propagators of (scalar)
light particles have the form  
\begin{eqnarray}
\label{Dc}
D_i(k)=\frac{1}{(2\pi)^4}\frac{1}{(m_i^2-k^2-i 0)}.
\end{eqnarray}
We neglect possible differences between the masses $m_i$ and use everywhere the same parameter $m$. 

Throughout the paper we work in the $B$-rest frame and for the sake of simplicity consider the case $q^2=q'^2=0$. 
We choose the $(+)$ and the $(-)$ directions of the light cone as follows:\footnote{
For any 4-vector $a$ we use light-cone coordinates, $a\equiv(a_+,a_-,a_\perp)$,
$a^2=2a_+a_--a_\perp^2$, $ab=a_+b_-+a_-b_+-a_\perp b_\perp$.} 
\begin{eqnarray}
\label{refframe}
q=(q_+,0,0),\quad q'=(0,q'_-,0),\quad p=(q_+,q'_-,0),\quad q_+=q'_-=M_B/\sqrt{2}, 
\end{eqnarray}
such that both $q_+$ and $q'_-$ are large.  

Let us derive the leading-order behaviour of the amplitude $A(q,q')$ for large $m_b$. 
\subsection{The $x$-vertex}
We start with the $x$-vertex and introduce $l=k+q$, the momentum carried by the light constituent field $\phi(x)$: 
\begin{eqnarray}
 \int \frac{dx d l}{m^2-(l-q)^2}e^{i l x}
 \langle 0|\phi(x)...|B(p)\rangle.
 \nonumber
\end{eqnarray}
Almost full momentum of the heavy $B$-meson is carried by the heavy $b$-quark; 
all light constituents carry only $O(\Lambda_{\rm QCD}/M_B)$ fraction of the $B$-meson momentum,
and thus the momentum $l$ is soft, $l_\mu\sim O(\Lambda_{\rm QCD})$. This feature is provided by
the properties of the $B$-meson wave functions. 

The component $q_+\sim M_B$ is large, and the propagator is highly virtual,
\begin{eqnarray}
  \label{2.5}
m^2-2 l_-(l_+-q_+)+l_\perp^2\sim \Lambda_{\rm QCD}M_B.
\end{eqnarray}
Let us expand the field operator $\phi(x)$ near $x=0$. The expansion in powers of $x_-$ and $x_\perp$ leads
to a well behaved Taylor series for the amplitude $A(q,q')$ because
\begin{eqnarray}
  x_{-}e^{i l_{+}x_{-}}
  \frac{1}{m^2-2l_-(l_+-q_+)+l_\perp^2}
\to e^{i l_+ x_-}\frac{l_-}{(m^2-2l_-(l_+-q_+)+l_\perp^2)^2}.
\end{eqnarray}
Since $l_-=O(\Lambda_{\rm QCD})$ and the virtuality of the propagator is $O(\Lambda_{\rm QCD}M_B)$, 
the contribution to $A(q,q')$ coming from any term $(x_-)^n$ is suppressed by a factor $(1/M_B)^n$
compared to the contribution of the term $(x_-)^0$. The same property holds for $(x_\perp)^n$.

However, for powers of the variable $x_+$ the situation is different: 
\begin{eqnarray}
  x_+ e^{i l_- x_+}\frac{1}{m^2-2l_-(l_+-q_+)+l_\perp^2}
  \to e^{i l_- x_+}\frac{q_+}{(m^2-2l_-(l_+-q_+)+l_\perp^2)^2}.
\nonumber
\end{eqnarray}
Since $q_+\sim M_B$, the contribution to $A(q,q')$ coming from any power of $x_+^n$, including  $x_+^0$,
has the same order of magnitude: the Taylor expansion of $\phi(x_+)$ near $x_+=0$ leads to no hierarchy
in the induced expansion of $A(q,q')$. We therefore need to keep the full $x_+$-dependence of
the operator $\phi(x_+)$ on the light cone ($x^2=0$). The leading term of the expansion of $A(q,q')$
related to the $x$-vertex corresponds to expanding $\phi(x)$ near $x_-=0$, $x_\perp=0$ and reads 
\begin{eqnarray}
  \int \frac{dx_+ dx_-dx_\perp  dl_+ dl_- dl_\perp}{(2\pi)^4}
  \frac{1}{m^2-2(l_+-q_+)l_-+l_\perp^2}e^{il_+x_-+i l_{-}x_+-il_\perp x_\perp}
\langle 0|...\phi(x_+)...|B(p)\rangle.
\end{eqnarray}
The $x_-$ and $x_\perp$ integrals may be now taken and lead to $\delta(l_\perp)\delta(l_+)$.
Integrating these $\delta$-functions and keeping only large $\propto m_b$ terms in the denominator, we obtain: 
\begin{eqnarray}
\int \frac{dx_+ dl_-}{2\pi}\frac{1}{2 q_+ l_-}e^{i l_-x_+}
  \langle 0|...\phi(x_+)...|B(p)\rangle.
\end{eqnarray}

\subsection{The $x'$-vertex}

Now, $q'_-\sim M_B$ is the only nonzero and large component of the vector $q'$. The propagator has the form
\begin{eqnarray}
  \label{2.9}
m^2-2l'_-(l'_--q'_-)+{l'_\perp}^2\sim \Lambda_{\rm QCD}M_B.
\end{eqnarray}
Obviously, we can perform the Taylor expansion
of $\phi(x')$ near $x'_+=0$ and $x'_\perp=0$ but have to keep its full dependence on the variable $x_-$.
Taking this feature into account, for the dominant contribution of the $x'$-vertex we obtain 
\begin{eqnarray}
  \int dx'_- dl'_+\frac{1}{2 q'_- l'_+}e^{i l'_+x'_-}
  \langle 0|...\phi(x'_-)...|B(p)\rangle.
  \nonumber
\end{eqnarray}

\subsection{The 3-particle contribution to the amplitude}
Making use of the leading contributions of the $x$- and $x'$-vertices,
we obtain the following expression for the leading part of $A(q,q')$
[recall that we work in the reference frame (\ref{refframe})]:
\begin{eqnarray}
  \label{Apqfinal}
  A(q,q')=
  \int \frac{dx_+ dl_-}{2\pi}\frac{1}{2 q_+ l_-}e^{i l_-x_+}
  \int \frac{dx'_- dl'_+}{2\pi}\frac{1}{2 q'_- l'_+}e^{i l'_+x'_-}
  \langle 0|\phi(x_+)\phi_b(0)\phi(x'_-)|B(p)\rangle.
\end{eqnarray}
This is the desired factorization formula for the simplest $B$-meson decay amplitude 
of the generic weak form factor topology:
For large $m_b$, the amplitude is given by the convolution of the hard kernel composed of 
highly virtual propagators of light degrees of freedom and the 3-particle wave function in the following
noncollinear LC configuration: the upper (above $b$-quark line) and the lower 
(below $b$-quark line) parts of the diagram are aligned along different LC directions.
We shall prove in the next Sections that this property holds also for diagrams
with the insertions of arbitrary numbers of light fields in the propagator line.

\newpage

\begin{figure}[t]
  \begin{tabular}{cc}
    \includegraphics[height=5cm]{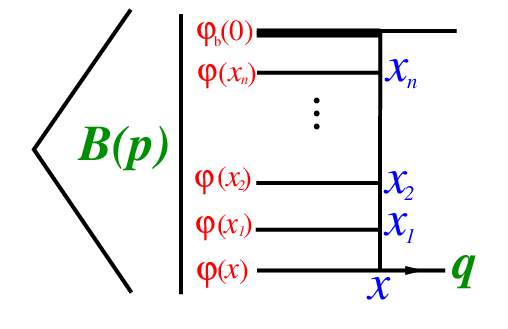}\hspace{1cm} & \hspace{1cm}
    \includegraphics[height=5cm]{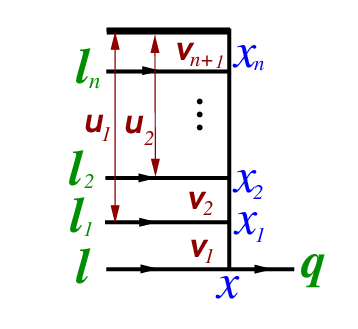}\\
    (a) \hspace{1cm}  & \hspace{1cm}  (b)
  \end{tabular}
  \caption{\label{Fig:2}
    (a) A typical diagram of the SL topology: the heavy field $\phi_b(0)$ and the light field $\phi(x)$
    hit, respectively, the upper and the lower end points of the vertical
    line along which light degrees of freedom propagate, with $n$ intermediate points corresponding
    to $n$ light fields $\phi(x_1)...\phi(x_n)$. The fields $\phi(x)$, $\phi(x_1),...,\phi(x_n)$ as
    well as the propagating light degrees of freedom may be different (e.g. light quarks and gluons). 
    (b) Coordinate, momentum and Feynman parameter notations for the amplitude of figure (a).}
\end{figure}
\section{\label{Sec:3}Multiparticle contribution to the amplitude of semileptonic topology}
Let us consider the amplitude of the type shown in Fig.\ref{Fig:0a}, where the heavy field $\phi_b(0)$
hits the end-point of the line along which energetic light degrees of freedom propagate, and the fields
$\phi(x),\phi(x_1),...,\phi(x_n)$ correspond to light particles.\footnote{Let us
emphasize that the light fields may be identical or may be different.
We do not indicate the type of the field by a separate index $\phi_i(x_i)$ just to make the notations more readable.
A typical configuration is when the fields $\phi(x_1),...,\phi(x_n)$ are gluons and the
field $\phi(x)$ is the light-quark field which also propagates along the vertical line
in Fig.~\ref{Fig:0a}.} We refer to this topology as to SL form factor topology. The amplitude has the form
\begin{eqnarray}
\label{ASL}
A^{(n)}_{\rm SL}(q|p)&=&\int dl\, dl_1...dl_n\,dx\, dx_1...dx_n\,e^{i(lx +l_1x_1+...+l_nx_n)}\nonumber\\
&&\times D(q-l)\cdots D(q-l-...-l_n)
\Psi(x,x_1,...,x_n,0|p).
\end{eqnarray}
Here the $(n+2)$-particle wave function is defined according to 
\begin{eqnarray}
 \Psi(x,x_1,...,x_n,0|p)=
\langle 0|\phi(x)\phi(x_1)...\phi(x_n)\phi_b(0)|B(p)\rangle. 
\end{eqnarray}
The propagators are given in Eq.~(\ref{Dc}).

Let us consider the amplitude (\ref{ASL}) in the $B$-meson rest frame.
In this reference frame, the momentum $q$ has at least one large LC component,
either $q_+\sim M_B$ or $q_-\sim M_B$, whereas all components of the vectors $l$ are soft,
$l_\mu\sim \Lambda_{\rm QCD}$.
Recall that this feature is provided by the $B$-meson wave function and reflects the fact that the heavy $b$-quark carries
almost the full momentum of the heavy $B$-meson, whereas each light valence degree of freedom
carries $\sim\Lambda_{\rm QCD}/M_B$ fraction of the $B$-meson momentum.
As the result, all propagators in the amplitude (\ref{ASL}) are highly virtual,
$k_i^2\sim m_b \Lambda_{\rm QCD}$.

\subsection{The propagators}
The amplitude contains $n$ intermediate points $x_1,\dots, x_n$ and $(n+1)$ propagators.
The standard procedure to handle these propagators is to introduce $(n+1)$ Feynman parameters $v_1,...,v_{n+1}$
satisfying $v_1+...+v_{n+1}=1$, $0<v_i<1$ to gather $(n+1)$ propagators into
the $(n+1)$-th power of the inverse single quadratic polynomial 
\begin{eqnarray}
P_2(k,l_1,...,l_n)=k^2 v_1+(k-l_1)^2v_2+...+(k-l_1-...-l_n)^2v_{n+1},\quad k=q-l, 
\end{eqnarray}
where mass terms have been omitted. We can write 
\begin{eqnarray}
P_2(k,l_1,...,l_n)&=&k^2\underbrace{(v_1+v_2+...+v_{n+1})}_{=1}
  -2\,k\, l_1\underbrace{(v_2+...+v_{n+1})}_{1-v_1\equiv u_1}
  -...-2\,k\, l_n\underbrace{(v_{n+1})}_{v_{n+1}\equiv u_n}
  +O(l_1,l_2,...,l_n)^2\nonumber\\
  &=&
  (k-u_1 l_1-u_2 l_2-\dots-u_n l_n)^2+O(l_1,l_2,...,l_n)^2=(q-\tilde l)^2+O(l_1,l_2,...,l_n)^2, 
\end{eqnarray}
where $O(l_1,l_2,...,l_n)^2$ denotes a quadratic polynomial composed of vectors $l_1,...,l_n$ and 
\begin{eqnarray}
\label{tildel}
\tilde l\equiv  l+u_1 l_1+u_2 l_2+\dots+u_n l_n),\qquad  0<u_n<u_{n-1}...<u_2<u_1. 
\end{eqnarray}
Since all components of the 4-vectors $l_1,...,l_n$ are soft,
and $(q-\tilde l)^2\sim O(m_b\Lambda_{\rm QCD})$,
we can expand the denominator in powers of small parameters
$l_il_j/(q-\tilde l)^2\simeq\Lambda_{\rm QCD}/m_b$. 
The leading contribution to the amplitude (\ref{ASL}) 
corresponds to setting $l_1,...,l_n\to 0$ in the denominator, or, equivalently,
to replacing $P_2(k,l_1,...,l_n)\to (q-\tilde l)^2$. 

\subsection{The exponential and the multiparticle wave function}
Let us make the following change of variables $(x_1,...,x_n)\to (z_1,...,z_n)$
\begin{eqnarray}
  x_i=u_i x+z_i,\quad i=1,...,n. 
\end{eqnarray}
The valiables $z_1,...,z_n$ measure the deviations of $x_1,...,x_n$ from
the straight line joining the points 0 and x. 

Then the exponential takes the form 
\begin{eqnarray}
\exp(ixl+i x_1l_1+...+ix_n l_n)=\exp(ix \tilde l+iz_1 l_1+...+i z_n l_n), 
\end{eqnarray}
with $\tilde l$ given in (\ref{tildel}). The wave function in the new variables reads 
\begin{eqnarray}
  \Psi(x,x_1,...,x_n,0|p)\to \Psi(x,u_1x_1+z_1,...,u_n x+z_n,0|p)
\end{eqnarray}
\subsection{The amplitude}
We are ready to obtain the leading contribution to the SL amplitude.
As discussed above, we can set $l_1,...,l_n\to 0$ in the denominator, coming to  
\begin{eqnarray}
A&=&\int D(u_1,...,u_n) 
\int d\tilde l dl_1...dl_n\int dx dz_1... dz_n\frac{1}{(2\pi)^{4(n+1)}}
  \nonumber\\
&&\times \frac{e^{ix \tilde l+i z_1l_1+...+i z_n l_n}}{(2q\tilde l)^{n+1}}
  \langle 0|\phi(x)\phi_1(u_1 x+z_1)...\phi_n(u_n x+z_n)\phi_b(0)|B(p)\rangle
\end{eqnarray}
with
\begin{eqnarray}
\int  D(u_1,...,u_n)(...)\equiv n! \int\limits_0^1 du_1\int\limits_0^{u_1}du_2...\int\limits_0^{u_{n-1}}du_n(...). 
\end{eqnarray}
The $l_1,...,l_n$ integrals may be taken leading to
$\int dl_i \exp(i l_i z_i)=(2\pi)^4\delta(z_i)$. Further, integrating over $dz_1...dz_n$, gives
\begin{eqnarray}
  A(q,q')&=&\frac{1}{(2\pi)^4}\int D(u_1,...,u_n) 
\int dx\,d\tilde l \frac{e^{ix \tilde l}}{(2q\tilde l)^{n+1}}
\langle 0|\phi(x)\phi_1(u_1 x)...\phi_n(u_n x)\phi_b(0)|B(p)\rangle. 
\end{eqnarray}
We work in the $B$-meson rest frame, $p=(p_+,p_-,p_\perp)=(M_B/\sqrt{2},M_B/\sqrt{2},0)$. 
In this reference frame, it is convenient to take $q$
along the $(+)$ or along the $(-)$ axis on the LC.

As we have seen in Sec.~\ref{Sec:2}, choosing $q$ in the form $q=(q_+,0,0)$, $q_+=M_B/\sqrt{2}$, we
can expand the arguments of the operators $\phi(u_i x)$ in powers of $x_-$ and $x_\perp$
near $x_-=0$ and $x_\perp=0$, integrate over $l_+,l_\perp$, and further over $x_-,x_\perp$,
coming to the expression 
\begin{eqnarray}
\label{ASLplus}
A_{\rm SL}(q_+)&=&\int D(u_1,...,u_n) 
\frac{1}{2\pi}
\int  dx_+\,d\tilde l_- \frac{e^{ix_+ \tilde l_-}}{(2q_+\tilde l_-)^{n+1}}
\langle 0|\phi(x_+)\phi_1(u_1 x_+)...\phi_n(u_n x_+)\phi_b(0)|B(p)\rangle. 
\end{eqnarray}
But we have also another possibility: if we choose $q$ along the $(-)$ LC axis,
$q=(0,q_-,0)$, $q_-=M_B/\sqrt{2}$, we obtain
\begin{eqnarray}
\label{ASLminus}
A_{\rm SL}(q_-)&=&\int D(u_1,...,u_n) 
\frac{1}{2\pi}
\int dx_-\,d\tilde l_+ \frac{e^{ix_- \tilde l_+}}{(2q_-\tilde l_+)^{n+1}}
\langle 0|\phi(x_-)\phi_1(u_1 x_-)...\phi_n(u_n x_-)\phi_b(0)|B(p)\rangle. 
\end{eqnarray}
Both representation are equivalent to each other and give the leading contribution to
$A_{\rm SL}$ valid up to $\Lambda_{\rm QCD}/m_b$ corrections. 

In the next Section we make use of both representation to obtain the leading 
behaviour of the form factor of the generic topology when the
heavy-quark field hits an intermediate point of the propagator line.  


\section{\label{Sec:4} Multiparticle contribution to the amplitude of the generic form factor topology}
We now generalize the results of the previous Sections to the case of diagrams of the
generic form factor topology shown in Fig.~\ref{Fig:0a}:
\begin{eqnarray}
\label{AFF}
A^{(n,n')}(q,q')&=&
\int dl\, dl_1...dl_n\,dx\, dx_1...dx_n\,e^{i(lx +l_1x_1+...+l_nx_n)}
\int dl'\, dl'_1...dl'_{n'}\,dx'\, dx'_1...dx'_{n'}\,e^{i(l'x' +l'_1x'_1+...+l'_{n'}x_{n'})}
\nonumber\\
&&\times D(q-l)\cdots D(q-l-...-l_n)D(q'-l')\cdots D(q-l'-...-l_{n'})
\nonumber\\
&&\times
\Psi(x,x_1,...,x_n,0,x'_{n'},...,x'_1,x'|p).
\end{eqnarray}
with the multiparticle wave function of the $B$-meson defined according to 
\begin{eqnarray}
 \Psi(x,x_1,...,x_n,0,x'_{n'},...,x'_1,x'|P)=
 \langle 0|\phi(x)\phi(x_1)...\phi(x_n)\phi_b(0)\phi(x'_{n'})...\phi(x'_1)\phi(x')|B(p)\rangle. 
\end{eqnarray}
To obtain factorization formula for $A^{(n,n')}(q,q')$, we again work in the $B$-rest frame
specified by Eq.~(\ref{refframe}). We can consider separately the
{\it upper part} of the diagram in Fig.\ref{Fig:0a} (i.e., above the point $x=0$ where $b$-quark hits the line) and
the {\it lower part} of the diagram (i.e., below the point $x=0$ where $b$-quark hits the propagator line).
The procedure is obvious:

\noindent
(i) For the lower part of the diagram, where $q_+$ emitted from the end-point $x$, has the large $(+)$-component,
we repeat the arguments leading to the factorization formula for $A_{\rm SL}(q_+)$, Eq.~(\ref{ASLplus}). 
We take into account that the diagram involves $n$ insertions of the light fields $\phi(x_1)...\phi(x_{n})$
in the propagator line.

\noindent
(ii) For the upper part of the diagram, where $q'$, emitted from the end-point $x'$, has the large
$(-)$-component, we repeat the arguments leading to the factorization formula for $A_{\rm SL}(q'_-)$,
Eq.~(\ref{ASLminus}). We take into account that the diagram involves $n'$ insertions
of the light fields $\phi(x'_{n'})...\phi(x'_{1})$ in the propagator line.

The dominant contribution to $A^{(n,n')}(q,q')$ is then found to be given by the
following convolution formula: 
\begin{eqnarray}
  \label{Annprime}
  A^{(n,n')}(q_+,q'_-)&=&
  \int D(u_1,...,u_n) \int \frac{dx_+\,d\tilde l_-}{2\pi}
\frac{e^{ix_+ \tilde l_-}}{(2q_+\tilde l_-)^{n+1}}
\int D(u'_1,...,u'_n)\int \frac{dx'_-\,d\tilde l'_+}{2\pi}
\frac{e^{ix'_- \tilde l'_+}}{(2q'_-\tilde l'_+)^{n'+1}}\nonumber\\
&&\times\Psi(x_+,u_1x_+,...,u_n x_+,0,u'_{n'}x'_-,...,u'_1x'_-,x_-'|p). 
\end{eqnarray}
This is a desired factorization theorem which represents the dominant contribution to the amplitude
of the generic form factor topology by the convolution of the hard kernel
(composed of hard propagators of light degrees of freedom) and the muliparticle wave function of the $B$-meson
$\Psi(x,x_1,...,x_n,0,x'_{n'},...,x'_1,x'|p)$ in the following {\it double-collinear} light-cone configuration
of Fig.~\ref{Fig:0b}:
\begin{eqnarray}
& x=(x_+,0,0),    \quad & x_1=(u_1 x_+ ,0,0),\quad \,\,\ldots,\quad     x_n=(u_n x_+,0,0),   \qquad 0<u_n<\ldots<u_1<1, \nonumber\\
&  x'=(0,x'_-,0), \quad & x'_1=(0, u'_1 x'_-,0),\quad \ldots,\quad x'_{n'}=(0,u'_{n'}x'_-,0), \,\quad 0<u'_{n'}<\ldots<u'_1<1.
  \nonumber
\end{eqnarray}
Here, the coordinates $x,x_1,\ldots,x_n$ are ordered and lie on the $(+)$-axis of the LC, 
and the coordinates $x',x'_1,\ldots,x'_{n'}$ are ordered and lie on the $(-)$-axis of the LC. 
That is why we refer to this configuration as to the {\it double collinear} LC configuration.

\newpage
\section{\label{Sec:5}Remark on the radiative corrections}

We have proven factorization formula to the leading order in $1/m_B$ and to zero order in $\alpha_s$.
[To be more precise, no loop corrections to the hard kernel have been considered].
However, a complete analysis of the factorization properties of the amplitude $A^{(n,n')}(q,q')$ requires 
the analysis of the radiative corrections. This is far beyond the scope of our discussion,
so we just outline here the main problems to be addressed. 

To zero order $\alpha_s$, the amplitude of Fig.~\ref{Fig:0a} splits into two pieces:
The upper part of the diagram (above the heavy-quark line) is aligned and ordered along say the
$(+)$-direction of the LC, then the lower part (below the heavy-quark line) is aligned and ordered
along the $(-)$-direction of the LC. 

We now want to consider the radiative corrections. They may be classified in the following way:
\begin{itemize}
\item[(I.a)]
Radiative corrections with gluon exchanges between the lines within the upper part of the diagram of Fig.~\ref{Fig:0a}.

\item[(I.b)]
Radiative corrections with gluon exchanges between the lines within the lower part of the diagram of Fig.~\ref{Fig:0a}.

\item[(II)]
Radiative corrections by gluons connecting some lines of the upper part with the
lines of the lower part of the diagram of Fig.~\ref{Fig:0a}. 
\end{itemize}
Classes I and II should be discussed separately. 

\subsection{The corrections of the types I.a and I.b}
For the radiative corrections of these classes all the results for factorization
properties proven for the usual SL topology remain valid: For instance,
the property that the infrared (IR) divergences may be systematically absorbed into the
generalized wave functions leaving the hard kernels free of infrared (IR) divergences and
the resummation of the Sudakov double logs to all orders \cite{Korchemsky:1999qb}; 
the generalization of the factorizaiton formula to all orders in $\alpha_s$
\cite{Descotes-Genon:2002crx,Lunghi:2002ju,Bosch:2003fc}; 
a proof that the contribution to the amplitude of a SL topology coming from 
three-particle generalized wafe functions is free of the end-point singularities,
as soon as renormalization group analysis is performed \cite{Huang:2023jdu}. 
For the corresponding results an extensive literature exists, so we
refer to selected papers and references given
therein \cite{Beneke:2011nf,Wang:2016qii,Wang:2021yrr,Huang:2023jdu}. 

\subsection{The corrections of the type II}
The gluons are exchanged between the lines of the upper and the lower parts of the diagram:
To understand the behaviour of such diagrams, we recall that the light degrees of freedom in the
upper part of the diagram are fast and aligned along the $(+)$-axis, while the light degrees of
freedom are fast and aligned along the $(-)$-axis. Therefore, in the major part of the phase space,
the gluon exchanged between these lines are hard and have the virtuality $k^2\sim m_B^2$.
However, according to soft-collinear effective theory \cite{Bauer:2000yr}, the exchange of soft gluons between
quarks propagating along opposite light-cone directions gives rise to a soft function, which
incorporates Wilson lines in both directions, see e.g. \cite{Beneke:2019slt} for the corresponding formalism. 
The convolution of this soft function with the hard kernel is of leading order and should be
properly taken into account. 

On the basis of the existing results for quantities which represent some fragments
of a complicated amplitude $A^{(n,n')}(q,q')$, it seems plausible to expect that the
IR divergences may be absorbed into properly determined soft wave functions leaving the
convolution of these wave functions with the hard kernel IR finite. In this respect, one expects that 
the factorization formula established in this paper will be properly modified but not destroyed.
However, a detailed analysis of the IR singularities in the radiative corrections to the amplitude 
$A^{(n,n')}(q,q')$ is a formidable task that requires also renormalization-group analysis of $n$-particle
soft wave functions. This task lies well beyong the scope of this paper.

\newpage

\section{\label{Sec:6} Summary}
\noindent
1. Our main result is the factorization formula, established at leading order in $\alpha_s$,
Eq.~(\ref{Annprime}), which represents $A^{(n,n')}(q,q')$, 
the $B$-decay amplitude of the generic weak form factor topology of Fig.~\ref{Fig:0a},
as the convolution of 
(i) the hard kernel composed of propagators of light degrees of freedom and
(ii) the multiparticle wave function of the $B$-meson in the
double-collinear LC configuration of Fig.~\ref{Fig:0b}:
In the reference frame chosen such that $q$ lies along the $(+)$-axis and $q'$ lies along the $(-)$-axis, 
the sets of the field coordinates $\{x_1,...,x_n\}$ and $\{x'_1,...,x'_{n'}\}$ are ordered along the
$(+)$ and the $(-)$ direction of the light cone, respectively. 

The double collinear configuration may be described in a covariant way:
Let us introduce 2 light-like vectors $a_\mu$ and $a'_\mu$, $a^2=a'^2=0$, $a'a=-2$ \cite{braun2017} such that 
\begin{eqnarray}
q_\mu \sim a_\mu, \quad  q'_\mu\sim a'_\mu. 
\end{eqnarray}
Then the double-collinear configuration is the following one: 
($x_\mu=z a_\mu$, $x'_\mu=z' a'_\mu$, $x^2=0$, $x'^2=0$, $x'x\ne 0$): 
\begin{eqnarray}
&& \langle 0|\phi(z a_\mu)\phi(u_1 z a_\mu)...\phi(u_n z a_\mu)\phi_b(0)
 \phi(u'_{n'}z'a'_\mu)...\phi(u'_1 z' a'_\mu)\phi(z' a'_\mu)|B(p)\rangle,\nonumber\\
 &&\qquad\qquad 0<u_{n}<...<u_1<1,\quad 0<u'_{n'}<...<u'_1<1. 
\end{eqnarray}
Obviously, the double collinear LC configuration is different
from the collinear LC configuration of \cite{braun2017}
and is not reduced to the latter, see \cite{m2023}. 

The case of quark and gluon fields of QCD is technically more involved than the case of scalar
fields discussed in detail in this paper. However, fermion propagators lead to the appearance of
polynomials in the numerators of the integrals studied here. So our conclusion about the
double-collinear mutiparticle LC configuration that provides the dominant contribution to the $B$-decay
amplitude of the generic weak form factor topology remains valid. 

\vspace{.5cm}
\noindent
2. We discussed in detail the case $q^2=q'^2=0$.
Let us briefly comment on the case $q^2\ne 0$, $q'^2\ne 0$ and $q^2,q'^2\simeq \Lambda_{\rm QCD}^2\ll M_B^2$.
In this case, the 4-vectors $q$ and $q'$ obtain small but nonzero ``second'' LC components:
\begin{eqnarray}
q&=&\frac{1}{\sqrt2}\left(M_B-\frac{q'^2}{M_B},\frac{q^2}{M_B},0\right),\nonumber\\
q'&=&\frac{1}{\sqrt2}\left(\frac{q'^2}{M_B},M_B-\frac{q^2}{M_B},0\right). 
\end{eqnarray}
Then, the propagator in Eq.~(\ref{2.5}) takes the form
\begin{eqnarray}
m^2-2(l_--q_-)(l_+-q_+)+l_\perp^2.
\end{eqnarray}
However, since $q_{-}\sim q^2/M_B$, the addition to the propagator is $O(q^2)$ and thus does not change
its leading behaviour to $\Lambda_{\rm QCD}/M_B$ accuracy:
\begin{eqnarray}
m^2-2(l_--q_-)(l_+-q_+)+l_\perp^2\to 2l_-q_+.
\end{eqnarray}
A similar argument applies to the propagator in Eq.~(\ref{2.9}) which to $\Lambda_{\rm QCD}/M_B$ accuracy takes the form 
\begin{eqnarray}
m^2-2(l'_+-q'_+)(l'_--q'_-)+l_\perp^2\to 2l'_+q'_-.
\end{eqnarray}
In the end, the factorization formulas (\ref{Apqfinal}), (\ref{ASLplus}), (\ref{ASLminus}),
and (\ref{Annprime}) do not change their form within $O(\Lambda_{\rm QCD}/M_B)$ accuracy. 

In summary, this paper presents the factorization formula for the
$B$-decay amplitude $A^{(n,n')}(q,q')$ valid at leading order in $\alpha_s$.
This result is an important first step in the analysis of factorization properties of 
a complicated quantity $A^{(n,n')}(q,q')$ but requires further analysis of the radiative
corrections to the hard kernel; the latter contain IR singularites which should be properly handled.  
A complete future analysis should include a detailed study of these IR singularities
and the renormalization-group properties of the multiparticle soft functions.

\acknowledgments
I have pleasure to thank Yu-Ming Wang for illuminating discussions related to the subject of this paper.  
The research was carried out within the framework of the program \emph{Particle Physics and Cosmology}
of the National Center for Physics and Mathematics. 


\begin{thebibliography}{100}

\bibitem{offen2007}
A.~Khodjamirian, T.~Mannel, and N.~Offen,
{\it Form-factors from light-cone sum rules with B-meson distribution amplitudes},
Phys.~Rev.~{\bf D75}, 054013 (2007).

\bibitem{wang2022a}
B.~Y.~Cui, Y.~K.~Huang, Y.~L.~Shen, C.~Wang and Y.-M.~Wang,
{\it Precision calculations of $B_{d,s}\to (\pi, K)$ decay form factors in soft-collinear effective theory}, 
JHEP \textbf{03}, 140 (2023). 

\bibitem{japan2001}
H.~Kawamura, J.~Kodaira, C.-F.~Qiao, and K.~Tanaka, 
{\it B-meson light cone distribution amplitudes in the heavy quark limit}, 
Phys.~Lett.~{\bf B523}, 111 (2001), Erratum: Phys.~Lett. {\bf B536}, 344 (2002).
\bibitem{Neubert}
M.~Beneke, G.~Buchalla, M.~Neubert, and C.~T.~Sachrajda,
{\it Penguins with Charm and Quark-Hadron Duality}, 
Eur.~Phys.~J. {\bf C61}, 439 (2009).
\bibitem{m2022}
D.~Melikhov,
{\it Nonfactorizable charming loops in FCNC B decay versus B-decay semileptonic form factors},
Phys.~Rev.~{\bf D106}, 054022 (2022).


\bibitem{voloshin}
M.~B.~Voloshin, {\it Large $O(m_c^{-2})$ nonperturbative correction
to the inclusive rate of the decay $B\to X_s\gamma$},  
Phys.~Lett. {\bf B397}, 275 (1997). 

\bibitem{ligeti}
Z.~Ligeti, L.~Randall, and M.~B.~Wise, 
{\it Comment on nonperturbative effects in $\bar B\to X_s\gamma$},
Phys.~Lett. {\bf B402}, 178 (1997). 

\bibitem{buchalla}
G.~Buchalla, G.~Isidori, S.~J.~Rey, 
{\it Corrections of order $\Lambda_{\rm QCD}^2/m_c^2$ to inclusive rare $B$ decays}, 
Nucl.~Phys. {\bf B511}, 594 (1998). 

\bibitem{khod1997}
A.~Khodjamirian, R.~Ruckl, G.~Stoll, and D.~Wyler, 
{\it QCD estimate of the long distance effect in $B\to K^*\gamma$}, 
Phys.~Lett. {\bf B402}, 167 (1997). 

\bibitem{zwicky1}
P.~Ball and R.~Zwicky, 	
{\it Time-dependent CP Asymmetry in $B\to K^*\gamma$ as a (Quasi) Null Test of the Standard Model}, 
Phys.~Lett. {\bf B642}, 478 (2006). 

\bibitem{zwicky2}
P.~Ball, G.~W.~Jones, and R.~Zwicky, 	
{\it $B\to V\gamma$ beyond QCD factorisation}, 
Phys.~Rev.~{\bf D75}, 054004 (2007). 

\bibitem{hidr}
A.~Khodjamirian, T.~Mannel, A.~Pivovarov, and Y.-M.~Wang, 
{\it Charm-loop effect in $B\to K^{(*)} l^+l^-$ and $B\to K^*\gamma$}, 
JHEP {\bf 09}, 089 (2010).

\bibitem{gubernari2020}
N.~Gubernari, D.~van Dyk, J.~Virto, 
{\it Non-local matrix elements in $B_{(s)}\to \{K^{(*)},\phi\}\ell^+\ell^-$},
JHEP {\bf 02}, 088 (2021).
\bibitem{gubernari2022} 
N.~Gubernari, M.~Reboud, D. van Dyk, and J.~Virto,
{\it Improved theory predictions and global analysis of exclusive
$b\to s\mu^+\mu^-$ processes},
JHEP {\bf 09}, 133 (2022). 

\bibitem{mk2018}
A.~Kozachuk and D.~Melikhov, 
{\it Revisiting nonfactorizable charm-loop effects in exclusive FCNC $B$ decays},
Phys.~Lett.~{\bf B786}, 378 (2018).
\bibitem{m2019}
D.~Melikhov,
{\it Charming loops in exclusive rare FCNC $B$-decays}, EPJ Web Conf. {\bf 222}, 01007 (2019).
\bibitem{m2023}
D.~Melikhov,
{\it Three-particle distribution in the B meson and charm-quark loops in FCNC B decays},
Phys.~Rev.~{\bf D108}, 034007 (2023).
\bibitem{wang2022}
Q.~Qin, Y.~L.~Shen, C.~Wang and Y.~M.~Wang,
{\it Deciphering the long-distance penguin contribution to $\bar B_{d, s} \to \gamma \gamma$ decays},
Phys.~Rev.~Lett. {\bf 131}  091902 (2023).
\bibitem{wang2023}
Y.-K. Huang, Y.~Ji, Y.-L.~Shen, C.~Wang, Y.-M.~Wang, X.-C.~Zhao, 
{\it Renormalization-Group Evolution for the Bottom-Meson Soft Function},
Phys.~Rev.~Lett. {\bf  133}, 171901 (2024).
\bibitem{bbm2023}
I.~Belov, A.~Berezhnoy, and D.~Melikhov,
{\it Charming-loop contributions in $B_s\to \gamma\gamma$ decay},
Phys.~Rev.~{\bf D108}, 094022 (2023).
\bibitem{bbm2024}
  I.~Belov, A.~Berezhnoy, and D.~Melikhov,
{\it Nonfactorizable charming-loop contribution to FCNC $B_s\to \gamma ll$ decay}, 
Phys.~Rev.~{\bf D109}, 114012 (2024). 

\bibitem{braun2017}
V.~Braun, Y.~Ji, and A.~Manashov,
{\it Higher-twist B-meson Distribution Amplitudes in HQET}",
JHEP {\bf 05}, 022 (2017).


\bibitem{Korchemsky:1999qb}
G.~P.~Korchemsky, D.~Pirjol and T.~M.~Yan,
{\it Radiative leptonic decays of B mesons in QCD}, 
Phys. Rev. D \textbf{61}, 114510 (2000). 

\bibitem{Descotes-Genon:2002crx}
S.~Descotes-Genon and C.~T.~Sachrajda,
{\it Factorization, the light cone distribution amplitude of the B meson
and the radiative decay $B\to \gamma l\nu$}, 
Nucl. Phys. B \textbf{650}, 356-390 (2003). 

\bibitem{Lunghi:2002ju}
E.~Lunghi, D.~Pirjol and D.~Wyler,
{\it Factorization in leptonic radiative $\to \gamma\nu$ decays}, 
Nucl. Phys. B \textbf{649}, 349-364 (2003). 

\bibitem{Bosch:2003fc}
S.~W.~Bosch, R.~J.~Hill, B.~O.~Lange and M.~Neubert,
{\it Factorization and Sudakov resummation in leptonic radiative B decay}, 
Phys. Rev. D \textbf{67}, 094014 (2003).

\bibitem{Beneke:2011nf}
M.~Beneke and J.~Rohrwild,
{\it $B$ meson distribution amplitude from $\to\gamma l\nu$}, 
Eur. Phys. J. C \textbf{71}, 1818 (2011).

\bibitem{Wang:2016qii}
Y.~M.~Wang,
{\it Factorization and dispersion relations for radiative leptonic $B$ decay}, 
JHEP \textbf{09}, 159 (2016).

\bibitem{Wang:2021yrr}
C.~Wang, Y.~M.~Wang and Y.~B.~Wei,
{\it QCD factorization for the four-body leptonic B-meson decays}, 
JHEP \textbf{02}, 141 (2022).

\bibitem{Huang:2023jdu}
Y.~K.~Huang, Y.~Ji, Y.~L.~Shen, C.~Wang, Y.~M.~Wang, and X.~C.~Zhao,
{\it Renormalization-Group Evolution for the Three-Particle B-Meson Soft Function}, 
Phys. Rev. Lett. \textbf{133}, no.17, 171901 (2024).

\bibitem{Bauer:2000yr}
C.~W.~Bauer, S.~Fleming, D.~Pirjol, and I.~W.~Stewart,
{\it An Effective field theory for collinear and soft gluons: Heavy to light decays}, 
Phys.~Rev.~D \textbf{63}, 114020 (2001). 


\bibitem{Beneke:2019slt}
M.~Beneke, C.~Bobeth, and R.~Szafron,
{\it Power-enhanced leading-logarithmic QED corrections to $B_q \to \mu^+\mu^-$}, 
JHEP \textbf{10}, 232 (2019) [erratum: JHEP \textbf{11}, 099 (2022)]. 
\end{thebibliography}
\end{document}